\documentclass[
  prl,
  reprint,
  amsmath,
  amssymb,
  twocolumn,
  nofootinbib,
  tightenlines,
  floatfix
  ]{revtex4-1}

\usepackage[utf8]{inputenc}
\usepackage[T1]{fontenc}

\usepackage{bm}
\usepackage{dsfont}
\usepackage{braket}
\usepackage[separate-uncertainty=true]{siunitx}
\usepackage[final]{hyperref}
\hypersetup{hidelinks}
\usepackage[caption=false]{subfig}
\usepackage{graphicx}

\newcommand{\vb}[1]{\bm{\mathrm{#1}}}
\newcommand{\tb}[1]{\bm{\mathrm{#1}}}
\DeclareMathOperator{\snabla}{\nabla_{\mathrm{S}}}
\DeclareMathOperator{\slaplace}{\Delta_{\mathrm{S}}}
\def\I{i}
\renewcommand{\Re}{\operatorname{Re}}

\begin{document}

\title{Controlling stability and transport of magnetic microswimmers by an external field}

\author{Fabian R. Koessel}
\email[]{fkoessel@uni-mainz.de}
\affiliation{Institute of Physics, Johannes Gutenberg-University, Staudingerweg 7-9, 55128 Mainz, Germany}

\author{Sara Jabbari-Farouji}
\email[]{sjabbari@uni-mainz.de}
\affiliation{Institute of Physics, Johannes Gutenberg-University, Staudingerweg 7-9, 55128 Mainz \& Germany, Kavli Institute for Theoretical Physics, University of California, Santa Barbara, California 93106, USA}

\date{\today}

\begin{abstract}
  We investigate the hydrodynamic stability and transport of magnetic microswimmers in an external field using a kinetic theory framework. Combining linear stability analysis and nonlinear 3D continuum simulations, we show that for sufficiently large activity and magnetic field strengths, a homogeneous polar steady state is unstable for both puller and pusher swimmers. This instability is caused by the amplification of anisotropic hydrodynamic interactions due to the external alignment and leads to a partial depolarization and a reduction of the average transport speed of the swimmers in the field direction. Notably, at higher field strengths a reentrant hydrodynamic stability emerges where the homogeneous polar state becomes stable and a transport efficiency identical to that of active particles without hydrodynamic interactions is restored.
\end{abstract}


\maketitle

Self-propulsion in conjunction with fluid-mediated interactions in active suspensions give rise to a wealth of collective phenomena that are very distinct from those found in passive systems at equilibrium~\cite{ramaswamy_mechanics_2010,marchetti_hydrodynamics_2013,elgeti_physics_2015, zottl_emergent_2016,cisneros_dynamics_2011}.
Some examples include hydrodynamic instabilities that lead to spatio-temporal pattern formation~\cite{aditi_simha_hydrodynamic_2002,saintillan_instabilities_2008, ezhilan_instabilities_2013}, active turbulence~\cite{wensink_meso-scale_2012, dombrowski_self-concentration_2004, dunkel_fluid_2013,collectivechemo} and unusual rheological properties~\cite{rafai_effective_2010,Lopez2015,Clement2016, vincenti_actuated_2017}.
Moreover, microswimmers exhibit new motility patterns in response to external stimuli such as
chemical signals~\cite{adler_chemotaxis_1966,theurkauff_dynamic_2012,collectivechemo},
light~\cite{Garcia2013,martin_photofocusing_2016},
gravitational~\cite{gyrotacticmodel1,ten_hagen_gravitaxis_2014,gyro_2017,wolff_sedimentation_2013,stark_swimming_2016}
and magnetic fields~\cite{spormann_unusual_1987,guell_hydrodynamic_1988,waisbord_destabilization_2016,collectivemag}.
The control and regulation of collective motion of microswimmers via an external field offers a promising route for their exploitation in high-tech applications such as micro-scale \emph{cargo transport}, \emph{targeted drug delivery}, and \emph{microfluidic devices}~\cite{martel_flagellated_2009, houle_magneto-aerotactic_2016,magswim7,cargodelivery}.

Presently, a theoretical understanding of collective behavior and transport of microswimmers in an external field is largely missing. Here, we put forward a kinetic theory for active suspensions that extends the previous kinetic models of active suspensions~\cite{aditi_simha_hydrodynamic_2002,saintillan_instabilities_2008} to include the effects of an external field. Our model is applicable to any active suspension driven by an aligning torque exerted by an external field. Examples include magnetotactic bacteria (MTB) carrying an intrinsic magnetic moment~\cite{blakemore_magnetotactic_1975,bazylinski_magnetosome_2004, frankel_magnetosomes_2009, Reufer2014,Rupprecht2016} and artificial magnetic swimmers~\cite{dreyfus_microscopic_2005,magswim2,magswim3,magswim4,magswim5,magswim6,ghosh_controlled_2009,Babel2016,Guzman2016,magswimRev2017} in an external magnetic field or bottom-heavy swimmers in a gravitational field~\cite{ten_hagen_gravitaxis_2014}. MTBs driven by a sufficiently strong magnetic field exhibit particularly intriguing patterns of collective behavior such as band formation~\cite{guell_hydrodynamic_1988,spormann_unusual_1987} and pearling instability under flow~\cite{waisbord_destabilization_2016}. Thus, we focus on the dynamics of active magnetic suspensions in a uniform magnetic field.

\begin{figure}[h]
  \includegraphics{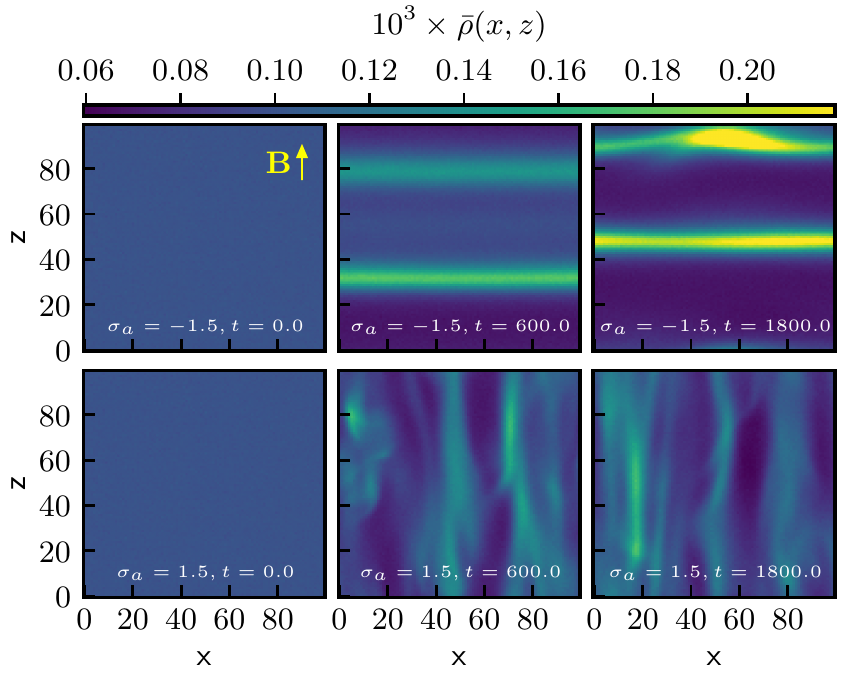}
  \caption[]{%
    Snapshots of density projections averaged along the y-axis from 3D non-linear simulations at different time steps (in dimensionless units) for pusher (top) and puller swimmers (bottom) in the unstable regime with dimensionless active stress \(|\sigma_a|=1.5\) and alignment parameter \(\alpha=4 \propto B/D_r\). The color encodes the probability density integrated in the y-direction \(\bar{\rho}(x, z) = \Delta y{\sum}_y \, \rho(x, y, z)\).
  }
  \label{fig:snapshots}
\end{figure}
We investigate the instabilities and transport of dilute suspensions of spherical magnetic swimmers in an external field combining linear stability analysis and 3D numerical simulations. We find that a homogeneous polar steady state is stable for low activity and magnetic field strengths but it becomes unstable at higher activity strengths and moderately strong magnetic fields for both pushers and pullers. These instabilities significantly reduce the polarization of the swimmers and lead to a decrease of their mean transport speed. In the unstable regime, we observe a rich phenomenology of pattern formation by varying the magnetic field and activity strengths. Representative examples of patterns for pushers and pullers are shown in Fig.~\ref{fig:snapshots}. Notably, pushers and pullers exhibit distinct instability patterns. The pushers concentrate in band-like structures perpendicular to the magnetic field that migrate in the direction of the magnetic field whereas pullers form lane-like patterns parallel to the field. Our results for the pushers are remarkably similar to the observed magnetotactic bands reported for spherical MTBs~\cite{guell_hydrodynamic_1988,spormann_unusual_1987}. The instability of the polar state induced by an external field shares similarities with the instability of aligned rod-like swimmers with nematic interactions~\cite{aditi_simha_hydrodynamic_2002,saintillan_instabilities_2008}. However, for an externally induced polar state such an instability disappears by further increase of field strength. To our knowledge such a \emph{reentrant hydrodynamic stability} has not been previously reported in active systems and calls for further experimental investigations.

\emph{Model system description.}--
We consider a dilute suspension of \( N \) spherical magnetic microswimmers with hydrodynamic radius \( a \) immersed in a fluid of volume \( V \) at a number density \(\varrho=\frac{N}{V}\). We assume that the self-propulsion is generated by a force-free mechanism of hydrodynamic origin such that the far field flow of a swimmer is well represented by that of a point-force dipole with an effective dipolar strength \(S_{\mathrm{eff}}\)~\cite{saintillan_instabilities_2008, ishikawa_suspension_2009,lauga_hydrodynamics_2009,adhyapak_flow_2017}.
\(S_{\mathrm{eff}}\) depends on the geometrical parameters of the model swimmer~\cite{magswim3,magswim4,magswim6,adhyapak_flow_2017},
for instance on \(a\) and the flagellum length \(\ell\)~\cite{adhyapak_flow_2017}. Each swimmer carries a weak magnetic dipole moment \(\mu_m\) along its body axis \(\vb{\hat{n}}\) and has a self-propulsion velocity \(v_{\mathrm{sp}}\vb{\hat{n}}\). The suspension is exposed to a uniform magnetic field \( \vb{B} \) that exerts an aligning torque on each swimmer. The magnetic moment values of MTBs are of the order of
\( \mu_m \approx \SI{1e-16}{\joule \per \tesla}\)~\cite{bazylinski_magnetosome_2004, meldrum_electron_1993,Fardin2013,Reufer2014} and their size \(a \sim \SI{1}{\micro \metre}\).
For dilute suspensions with inter-particle distances \(r \gtrsim 5 a\), their dipole-dipole interactions are small compared to the thermal energy scale and we can neglect their effect. Instead, we focus on the interplay between the hydrodynamic interactions and the aligning torque.

\emph{Kinetic theory.}--
For sufficiently low \(\varrho\), the mean-field configuration of an ensemble of the swimmers at a time \(t\) can be described by the probability density \( \frac{1}{V} \Psi( \vb{x}, \vb{\hat{n}}, t) \) of finding a particle with the center of mass position \(\vb{x}\) and the unit orientation \( \vb{\hat{n}} \). \( \Psi \) is normalized such that \( \frac{1}{V} \int d\vb{x} \int d\vb{\hat{n}} \Psi = 1 \). The time evolution of \( \Psi \) is governed by a \emph{Smoluchowski}-equation of the form:
\begin{equation}
  \partial_{t} \Psi
  + \nabla \cdot \vb{J}_{\mathrm{tr}}[\Psi]
  + \snabla \cdot \vb{J}_{\mathrm{rot}}[\Psi]
  =\mathbb{D} \Psi,
  \label{eq:fokkerplanck}
\end{equation}
where
\(
  \snabla = (\mathds{1} - \vb{\hat{n}} \vb{\hat{n}}) \cdot \nabla_{\vb{\hat{n}}}
\)
denotes the angular gradient operator; \(\vb{J}_{\mathrm{tr}}\) and \(\vb{J}_{\mathrm{rot}}\) are the translational and rotational drift currents. \( \mathbb{D} = D_t \Delta + D_r \slaplace \) accounts for the evolution of \( \Psi \) resulting from the translational and rotational diffusive currents. \( D_t \) and \( D_r \) represent the effective long-time translational and rotational diffusion coefficients that can be of thermal or biological origin, {\it e.g.} due to tumbling of bacteria.
\( \vb{J}_{\mathrm{tr}} \equiv \Psi \vb{v_{\mathrm{tr}}} \)
describes the translational current stemming from the self-propulsion of a swimmer and its convection in the local flow \(\vb{u}\),
\begin{equation}
  \vb{v}_{\mathrm{tr}} =
    v_{\mathrm{sp}} \vb{\hat{n}} + \vb{u}[\Psi].
  \label{eq:trans_flux_vel}
\end{equation}
The rotational current
\( \vb{J}_{\mathrm{rot}} \equiv \Psi \vb{v_{\mathrm{rot}}} \)
incorporates contributions from the rotational velocities resulting from the torque generated by the aligning magnetic field and the local flow vorticity
\(
  W_{jl} \equiv 1/2 ( \partial_j u_l - \partial_l u_j )
\)
according to the \emph{Jeffery}'s equation~\cite{jeffery_motion_1922,junk_new_2007}:
\begin{equation}
  \vb{v}_{\mathrm{rot}} =
    (\mathds{1} - \vb{\hat{n}} \vb{\hat{n}}) \cdot
    \left(
      \mu_m / \xi_{\mathrm{R}} \vb{B}
      - \tb{W}[\vb{u}] \cdot \vb{\hat{n}}
    \right),
  \label{eq:rot_flux_vel}
\end{equation}
where \( \xi_{\mathrm{R}} \) is the rotational friction coefficient.

The flow field \( \vb{u}[\Psi] \) in the low Reynolds number limit is captured by the incompressible \emph{Stokes} equation and is determined by the state of the system encoded by \(\Psi\) via a mean-field stress profile \(\tb{\Sigma}[\Psi]\). It includes two contributions: an active stress \(\tb{\Sigma}_a\), generated by the self-propulsion of force-free dipolar swimmers~\cite{ishikawa_suspension_2009,lauga_hydrodynamics_2009}, and an antisymmetric magnetic stress \( \tb{\Sigma}_m\), caused by reorientation of swimmers in the magnetic field. The active stress is proportional to the angular expectation value of the nematic order tensor
\(
  \tb{\Sigma}_a(\vb{x}) = \Sigma_a
  \int d\vb{\hat{n}} \Psi \,
     \left(\vb{\hat{n}} \vb{\hat{n}} - \frac{1}{3} \mathds{1} \right)
\)~\cite{doi_theory_2009,saintillan_instabilities_2008}.
The strength of the active stress is given by \(\Sigma_a = \pm \varrho S_{\mathrm{eff}}\). The sign of \(\Sigma_a\) determines the nature of the swimmers, being a puller \(\Sigma_a > 0\) or a pusher \(\Sigma_a < 0\). The magnetic stress is given by
\(
  \tb{\Sigma}_m(\vb{x}) = \frac{\Sigma_m}{2}
  \int d\vb{\hat{n}} \Psi \,
    \left( \vb{\hat{n}} \vb{\hat{B}} - \vb{\hat{B}} \vb{\hat{n}} \right),
\)
in which \( \vb{\hat{B}} = \vb{B} / B \) and \( \Sigma_m = \varrho \mu_m B \)~\cite{ilg_magnetization_2002}.

To facilitate the analysis of our model, we render the equations dimensionless, using the following characteristic velocity, length, and time scales: \(u_c=v_{\mathrm{sp}}\), \(x_c=\rho^{-1/3}\) (average inter-particle distance) and \(t_c=x_c/u_c\). These scaling choices leave the distribution function unchanged: \(
  \Psi(\vb{x}, \vb{\hat{n}}, t) \equiv
  \Psi_{\mathrm{scaled}}(\vb{x}/x_c, \vb{\hat{n}}, t/t_c)
\).
The corresponding dimensionless model parameters are the magnetic field strength
\(b = t_c \mu_m B / \xi_{\mathrm{R}}\),
the rotational and translational diffusion coefficients
\(d_r = D_r \varrho^{-1/3} v_{\mathrm{sp}}^{-1}\)
and
\(d_t = D_t \varrho^{1/3} v_{\mathrm{sp}}^{-1}\),
the active stress amplitude
\(
  \sigma_a
    = t_c \eta^{-1} \Sigma_a
    = \varrho^{2/3} S_{\mathrm{eff}} v_{\mathrm{sp}}^{-1} \eta^{-1}
\)
and the magnetic stress amplitude
\(
  \sigma_m
    = t_c \eta^{-1} \Sigma_m
    = \varrho^{2/3} \mu_m B v_{\mathrm{sp}}^{-1} \eta^{-1}
\)
in which \(\eta\) denotes the fluid viscosity.

\emph{ Homogeneous steady state solution.}--
Let us first consider a solution \(\Psi_0(\vb{\hat{n}})\) of equation~\eqref{eq:fokkerplanck} satisfying,
\( \partial_t \Psi_0 = 0 \) and \( \nabla \Psi_0 = 0 \).
It is given by
\begin{equation}
  \Psi_0(\vb{\hat{n}}) = \frac{\alpha}{4 \pi \sinh \alpha} e^{\alpha \vb{\hat{n}} \cdot \vb{\hat{B}}}.
  \label{eq:fixpoint}
\end{equation}
in which \( \alpha = b/d_r \equiv \mu_m B/ (\xi_{\mathrm{R}} \, D_r) \) and it is identical to the steady state solutions obtained in~\cite{Rupprecht2016,waisbord_destabilization_2016,vincenti_actuated_2017}. We call \(\alpha\) the alignment parameter as it is equal to the ratio of two characteristic reorientation times; \(\alpha \equiv \tau_r / \tau_m\).
\(\tau_r = D_r^{-1}\) represents the average decorrelation time of the particle from its initial orientation and \(\tau_m = \xi_{\mathrm{R}} / \mu_m B\) describes the typical time a non-diffusive particle needs to align itself with the magnetic field. The \(\Psi_0(\vb{\hat{n}})\) in Eq.~\eqref{eq:fixpoint} corresponds to a homogeneous polar state with a polarization vector
\(
  \vb{p} \equiv {\left< \vb{\hat{n}}\right>}=p_0(\alpha) \vb{\hat{B}}
\), where
\(
  {\left< \bullet \right>} \equiv \frac{1}{V}
    \int d\vb{x} \int d\vb{\hat{n}}\, \Psi(\vb{x}, \vb{\hat{n}}, t) \bullet
\)
defines the expectation value with respect to \(\Psi\). The polarization magnitude is given by
\begin{equation}
  p_0(\alpha) = (-1 + \alpha \coth \alpha) / \alpha,
  \label{eq:polarization}
\end{equation}
Note that a full alignment is only achieved for \(\alpha \gg 1 \).

\emph{Linear stability analysis.}--
We investigate the linear stability of the homogeneous polar state by considering a small perturbation of the form
\(
  \Psi_0 + \varepsilon \, \tilde{\Psi}(\vb{k}, \vb{\hat{n}})
  e^{\I \vb{k} \cdot \vb{x} + \lambda t}
\).
The equation of motion linearized in \( 0 < \varepsilon \ll 1 \) can be expressed as an eigenvalue problem of the form
\(
  \mathbb{L} \tilde{\Psi} = \lambda \tilde{\Psi}
\),
where \( \mathbb{L}(\vb{k},\vb{\hat{n}},\vb{b},\Psi_0) \) is a linear differentio-integro-operator~\cite{Supp}.
To solve this problem spectrally, we expand the perturbation amplitude \(\tilde{\Psi}\) in the basis of spherical harmonics \(Y_l^h\), {\it i.e.}
\(
  \tilde{\Psi} = \sum_{l=0}^{\infty} \sum_{h=-l}^{l} \ket{Y_l^h} \braket{Y_l^h | \tilde{\psi}(\vb{k})}
\)
and reduce it into an algebraic system of equations for the harmonic amplitudes:
\(
  \sum_{j=0}^{\infty} \sum_{m=-j}^{j} \braket{Y_l^h | \mathds{L} | Y_j^m}
  \braket{Y_j^m | \tilde{\psi}(\vb{k})}
  = \lambda \braket{Y_l^h | \tilde{\psi}(\vb{k})}
\).
We solve this algebraic eigenvalue problem numerically by truncating the sum for sufficiently large \(j\) such that the convergence of the dominant eigenvalues are ensured. The external field breaks the rotational symmetry. Hence, the stability depends on the direction \(\hat{\vb{k}}\) of the perturbation wave vector with respect to the magnetic field direction, which can be characterized by a single angle
\(\Theta_{\mathrm{B}}\equiv \cos^{-1}(\hat{\vb{k}} \cdot \hat{\vb{B}})\).

\begin{figure}[t]
  \label{fig:growthrate_k}
  \includegraphics{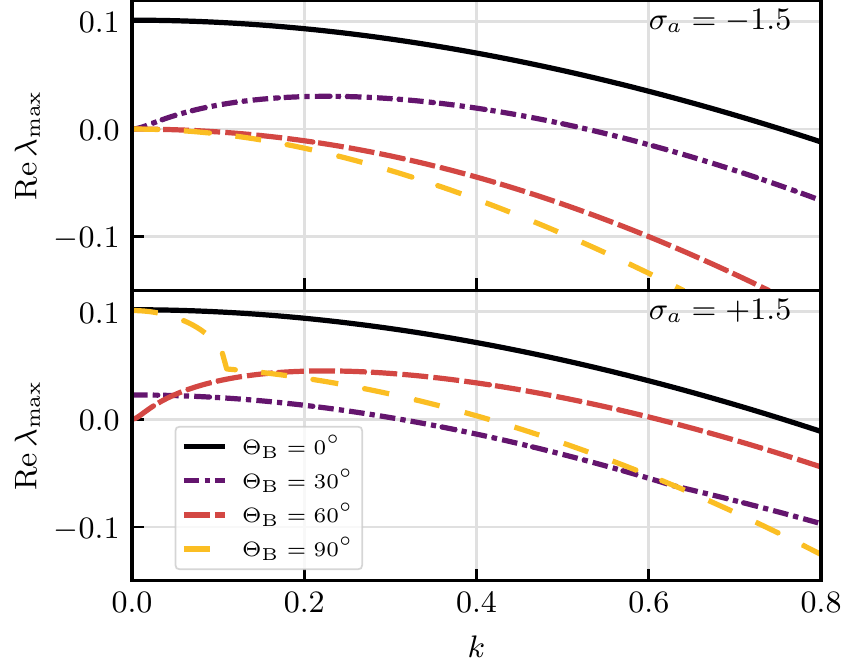}
  \caption{%
    The dependence of the largest growth rate \( \Re \, \lambda_\mathrm{max}\), on the wavenumber \(k\), for the homogeneous polar steady state given in Eq.~\eqref{eq:fixpoint} at several wave angles \(\Theta_\mathrm{B}\) for pushers (top, \(\sigma_a = -3/2\)) and pullers (bottom, \(\sigma_a = 3/2\)).
    The other dimensionless parameters are fixed to \(\alpha=4\), \(\sigma_m = 0.002\), \(d_r = 0.05\), and \(d_t = 0.1\).}
\end{figure}

We first examine the stability of swimmers with moderate values of activity and magnetic field strength leading to \(\sigma_a = \pm 3/2\) and \(\alpha=4\). The remaining parameters are chosen to be comparable to those of MTBs~\cite{waisbord_destabilization_2016} and they are fixed to: \(\sigma_m=0.01 \, \alpha \, d_r\), \(d_r = 0.05\), and \(d_t = 0.1\). Figure~\ref{fig:growthrate_k} shows the real part of the eigenvalue with the largest magnitude \( \Re \,\lambda_{\mathrm{max}}(k) \), the so-called maximum growth rate, as a function of \(k=|\vb{k}|\) at various perturbation angles \(\Theta_{\mathrm{B}}\).
For both puller and pusher swimmers, long wavelength perturbations dominate the instabilities.
For pushers, perturbations grow fastest in the direction of the magnetic field, whereas for pullers, perturbation directions parallel and perpendicular to \(\vb{B}\) predominate the instabilities. Thus, we expect pushers and pullers to exhibit distinct instability patterns as confirmed by the non-linear dynamics simulations; see Fig.~\ref{fig:snapshots}.

\begin{figure}[t]
  \label{fig:stability_sim}
  \includegraphics{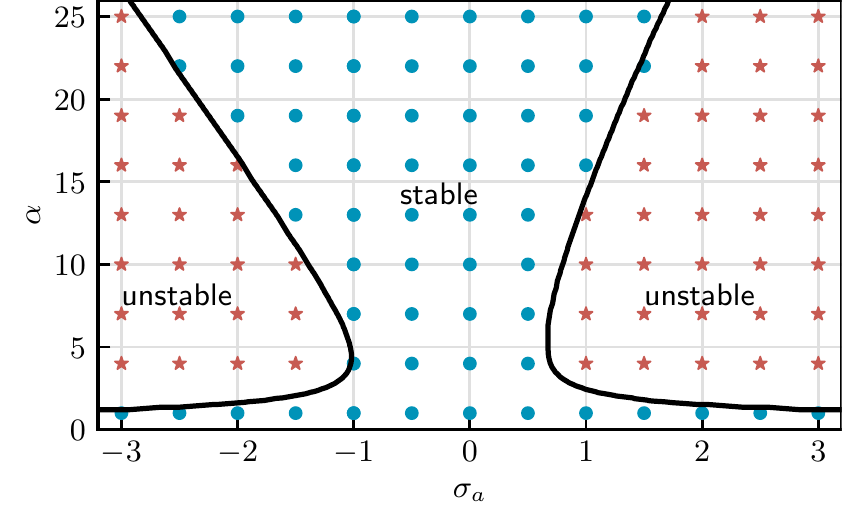}
  \caption{%
    Stability diagram of the homogeneous polar state given by Eq.~\eqref{eq:fixpoint} where we have varied the strengths of active stress \( \propto \sigma_a \) and magnetic field \( \propto \alpha \) assuming a constant volume fraction of
    \(\phi \approx 0.01\),
    \(a = \SI{1}{\micro \metre}\),
    and \(\mu_m = 10^{-16} J/T\).
    The dimensionless diffusion coefficients are fixed to \(d_r = 0.05\) and \(d_t = 0.1 \) and the reduced magnetic stress
    amplitude is varied as \(\sigma_m = 0.01 \, \alpha \, d_r\).
    The solid lines are determined using linear stability analysis and separate the stable from unstable regions. The circles and stars depict stable and unstable points in phase space found by non-linear dynamics simulations.
  }
\end{figure}

Having investigated the stability of the homogeneous polar state for fixed values of \(\alpha\) and \(\sigma_a\), we now present the stability phase diagram in which we vary the strengths of both activity \(S_{\mathrm{eff}} \propto \sigma_a\) and magnetic field \(B \propto \alpha\). Figure~\ref{fig:stability_sim} depicts the the stability diagram for \(\Psi_0(\alpha)\) in the (\(\sigma_a \), \(\alpha\)) plane. The magnetic stress is varied concomitant with \(\alpha\) as \(\sigma_m = 0.01 \alpha d_r \); the remaining parameters are kept constant at the values given in the caption. From the linear stability analysis, we determine the border lines that separate the stable from the unstable regions. The homogeneous polar state is only stable for low values of \(\sigma_a\) or \(\alpha\).
For larger \(\sigma_a\), as soon as a moderately strong magnetic field aligns the swimmers, the amplified anisotropic hydrodynamic interactions oppose the alignment in the \(\vb{B}\) direction. Consequently, the interplay between the hydrodynamic interactions and alignment torque gives rise to the instability of the steady state. Remarkably, for stronger magnetic fields the hydrodynamic instability can be overcome and the steady state becomes stable again. To examine the validity of these predictions, we study the dynamics of swimmers by non-linear simulations.

\emph{Non-linear dynamics simulations.}--
We perform nonlinear simulations of the kinetic model in three dimensions to study the long-time dynamics and pattern formation resulting from the instabilities. To solve the Smoluchowski equation Eq.~\eqref{eq:fokkerplanck} with periodic boundary conditions, we use a hybrid stochastic particle based sampling method to obtain \( \Psi(\vb{x},\vb{\hat{n}},t) \) and a spectral method to solve for the flow field \( \vb{u}(\vb{x}) \). For every time step, we integrate the corresponding Langevin stochastic differential equations for the positions and orientations of a large number (\(10^6 - 10^8\)) of independent and randomly initialized test particles. The test particle configurations provide us with sufficient statistics to construct a normalized histogram for spatial-orientational realization of \(\Psi(\vb{x},\vb{\hat{n}},t)\) from which we compute the stress profile in the fluid. Given the stress, we solve the Stokes equation for the flow field by expanding it in terms of Fourier modes on a grid. Eventually, \(\vb{u}(\vb{x})\) is fed back into the next integration time step for the Langevin equations. We use a grid of 100 lattice points with box dimensions of \(100\, x_c\) for each of the spatial coordinates, and 24 and 16 points for the spherical polar and azimuthal orientational coordinates \(\theta\) and \(\phi\) in \(\vb{\hat{n}}(\theta, \phi)\).

\begin{figure}
  \label{fig:cp}
  \includegraphics{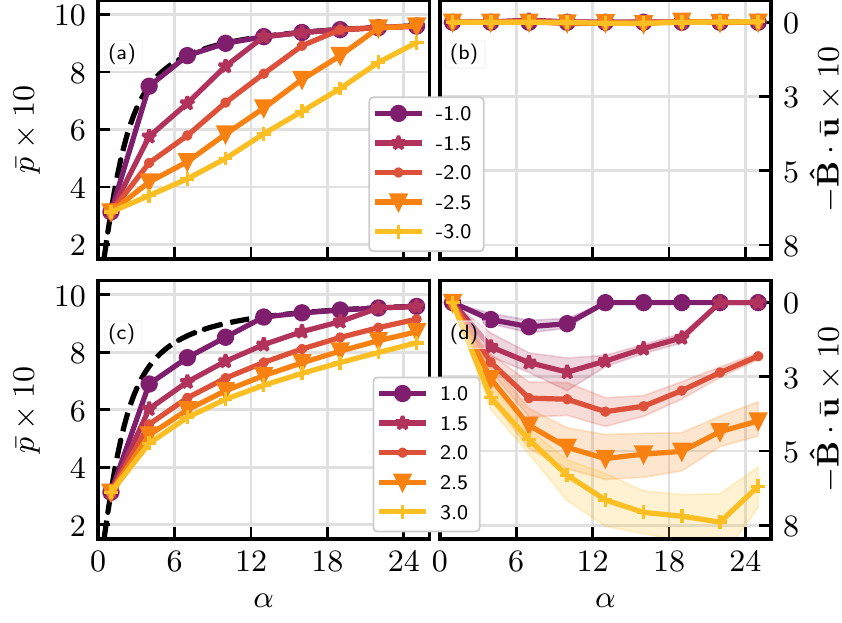}
  \caption{%
    (a), (b) The time-averaged polarization magnitude \(\bar{p}\) as a function of \(\alpha \propto B\) at different activity strengths \( \propto \sigma_a\) for pushers and pullers. The dashed line shows the polarization of the steady state given in Eq.~\eqref{eq:polarization}.
    (c), (d) The time-averaged convection transport speed in the magnetic field direction versus \(\alpha\) at different \(\sigma_a\) for pushers and pullers. The data are extracted from simulations with a box size of \(100^3 x_c^3\).
  }
\end{figure}

We explore the stability of \(\Psi_0(\alpha)\) for different activity and magnetic field strengths while keeping all the other parameters constant. The (\(\sigma_a,\alpha\)) values probed by non-linear simulations are shown by symbols in Fig.~\ref{fig:stability_sim}. For the points denoted by discs, \(\Psi\) evolves towards the homogeneous polar state given by Eq.~\eqref{eq:fixpoint}. Conversely, for the points depicted by stars, \(\Psi\) departs from \(\Psi_0(\alpha)\) and an inhomogeneous time-dependent density profile develops. Fig.~\ref{fig:stability_sim} demonstrates that the predictions of linear stability analysis and non-linear simulations for the unstable regions are in excellent agreement.
Density gradients in the inhomogeneous systems generate a flow with a non-zero vorticity field which is coupled to the swimmers orientation and rotates them away from the magnetic field direction. Thus, we expect that this effect results in a reduction of the average polarization. To confirm this hypothesis, we measure the time-averaged global polarization defined as
\(
  \bar{p} = \frac{1}{N_t} \sum_{j=0}^{N_t}
   \|\left< \vb{\hat{n}} \right>(t_0 + j \Delta t)\|
\),
where \(\Delta t\) is the simulation time step. \(t_0\) marks a relaxation time after which \(\bar{p}\) is nearly time-independent despite exhibiting non-stationary patterns~\cite{Supp}. The average polarization vector is almost parallel to the magnetic field; \(\vb{\hat{B}} \cdot \vb{\bar{p}} \approx \bar{p}\) and it is independent of the system size for \(L\gtrsim 50\)~\cite{Supp}.
Figs.~\ref{fig:cp}a and~\ref{fig:cp}b present the \(\bar{p}\) as a function of \(\alpha\) for pushers and pullers at different activity strengths \(\sigma_a\). For moderate \(\sigma_a\) and \(\alpha\) values falling in the unstable regime, we observe a significant reduction in the polarization compared to the the steady state polarization \(p_0(\alpha)\) (Eq.~\ref{eq:polarization}) shown by the dashed line. The decrease in polarization is stronger for larger activity strengths. Remarkably, stronger magnetic fields drive the system back into the stable regime and \(\bar{p}\) agrees with \(p_0(\alpha)\) in those regions.

The mean polarization determines the mean transport speed \(\bar{v}\) in the direction of magnetic field that additionally includes a contribution from the convective flow component along \(\vb{\hat{B}}\):
\begin{align}
  \bar{v}
    = \vb{\hat{B}} \cdot
      \left< v_{\mathrm{sp}} \vb{\hat{n}} + \vb{u} \right>
    = (v_{\mathrm{sp}} \vb{p} + \left< \vb{u} \right>) \cdot \vb{\hat{B}}.
  \label{eq:transportvelocity}
\end{align}
For an efficient transport in the direction of magnetic field, a high polarization of swimmers parallel to \(\vb{B}\) is desirable which can be achieved by increasing the magnetic field strength. To evaluate the contribution of the mean convective flow speed \(\left<\vb{u}\right>\) to the transport, we calculate the time-averaged mean value of flow field as
\(
  \vb{\bar{u}} = \frac{1}{N_t} \sum_{j=0}^{N_t}
    \left< \vb{u} \right>(t_0 + j \Delta t)
\).
Figs.~\ref{fig:cp}c and~\ref{fig:cp}d show \(\vb{\bar{u}} \cdot \vb{\hat{B}}\) versus \(\alpha \propto B\) for pushers and pullers at different values of \(\sigma_a\) that is almost independent of box size for \(L\ge 50\)~\cite{Supp}. The average flow created by pushers has a vanishing component along the magnetic field. Therefore, their average transport speed is governed by their mean polarization. By contrast, for the pullers the contribution of convective flow to the transport is not negligible. Pullers in the unstable regime concentrate in lane-like structures along \(\vb{\hat{B}}\) and predominantly create a convective flow component anti-parallel to the magnetic field that reduces the average transport speed along the magnetic field.
Fig.~\ref{fig:cp}d demonstrates that \(\vb{\bar{u}} \cdot \vb{\hat{B}}\) is a non-monotonic function of \(\alpha\).
Thus, an inefficient transport can be evaded by increasing the strength of \(\vb{B}\) and pushing the system towards the stable regime.

\emph{Conclusions.}--
Our results highlight the significance of hydrodynamic interactions in hindering the directed transport of swimmers in the unstable regime. We observe a novel reentrant hydrodynamic stability when increasing the field strength beyond an activity-dependent value. In the unstable regime, the magnetic suspensions exhibit distinct instability patterns for pusher and puller swimmers in the external field and proposes a pragmatic approach for distinguishing pushers from pullers in experiments. We defer a classification of patterns as a function of strengths of activity and magnetic field to a future work. Moreover, elucidating the role of swimmer-swimmer correlations~\cite{CorrPRL2017}, magnetic dipole-dipole and near-field hydrodynamic interactions in more concentrated suspensions merits further investigations.

\begin{acknowledgments}
  We thank Eric Clément, M. Cristina Marchetti and Friederike Schmid for fruitful discussions and Tapan Adhyapak for a critical reading of the manuscript. We acknowledge the financial support from the German Research Foundation (http://www.dfg.de) within SFB TRR 146 (https://trr146.de). The simulations were performed using the MOGON II computing cluster. This research was supported in part by the National Science Foundation under Grant No. NSF PHY17-48958.
\end{acknowledgments}

\bibliographystyle{apsrev4-1}
\bibliography{references}
\end{document}